\documentclass[11pt, a4paper]{article}
\pdfoutput=1

\usepackage{amsmath,amssymb}
\usepackage{mathrsfs,mathtools}
\usepackage{comment}
\usepackage{multirow}
\usepackage[utf8]{inputenc}
\usepackage{bm}
\usepackage{cite}
\usepackage{physics}
\usepackage{braket}
\usepackage{soul}
\usepackage[footnotesize]{caption}
\usepackage{acronym}

\usepackage{ifpdf}
\ifpdf        
  \usepackage{graphicx, hyperref, xcolor}     
\else     
  \usepackage[dvipdfmx]{graphicx, hyperref, xcolor}     
 \fi
\usepackage{subcaption}

\usepackage{tikz}
\usepackage{tikz-feynman}
\tikzfeynmanset{compat=1.0.0}
 
\definecolor{rossoferrari}{HTML}{D9073D}
\definecolor{mediumblue}{HTML}{0000CD}
\definecolor{forestgreen}{HTML}{228B22}
\definecolor{desy_blue}{HTML}{009EE2}
\definecolor{desy_orange}{HTML}{FD8800}
\hypersetup{
setpagesize=false,
bookmarksnumbered=true,%
bookmarksopen=true,%
colorlinks=true,%
linkcolor=rossoferrari,
urlcolor=mediumblue,
citecolor=mediumblue,
linktocpage=false,
}

\usepackage[height=8.85in,width=6.8in]{geometry}

\usepackage{fourier}

\makeatletter
\@addtoreset{equation}{section}

\makeatother

\def\mcA{\mathcal{A}}

\def\mcP{\mathcal{P}}

\def\mcW{\mathcal{W}}

\def\Mpl{M_{\rm Pl}}

\newcommand{\bmA}{\bm{A}}
\newcommand{\bmB}{\bm{B}}
\newcommand{\uc}{\mathrm{c}}
\newcommand{\bmE}{\bm{E}}
\newcommand{\bme}{\bm{e}}

\newcommand{\bmk}{\bm{k}}

\newcommand{\calP}{\mathcal{P}}
\newcommand{\up}{\mathrm{p}}
\newcommand{\bmx}{\bm{x}}
\newcommand{\bmy}{\bm{y}}

\newcommand{\Uone}{\mathrm{U}(1)}

\newcommand{\IR}{\mathrm{IR}}

\acrodef{EoM}{equation of motion}
\acrodef{dS}{de Sitter}
\acrodef{CS}{Chern--Simons}

\newcommand{\bae}[1]{\begin{align} #1 \end{align}}

\begin{document}

\begin{titlepage}

\begin{flushright}
KEK-TH-2434
\end{flushright}

\begin{center}


\vskip 1.in

{\Huge \bfseries
Stochastic formalism for $\Uone$ gauge fields
\\
in axion inflation
\\
}

\vskip .8in

{\Large
Tomohiro Fujita$^{\triangledown}$,
Kyohei Mukaida$^{\lozenge,\blacklozenge}$,
Yuichiro Tada$^{\bigstar,\lozenge}$
}

\vskip .3in

\begin{tabular}{ll}
$^\triangledown$&\!\!\!\!\!\! \emph{Waseda Institute for Advanced Study, Shinjuku, Tokyo 169-8050, Japan}\\[.3em]
$^\triangledown$&\!\!\!\!\!\! \emph{Research Center for the Early Universe, University of Tokyo, Bunkyo, Tokyo 113-0033, Japan}\\[.3em]
$^\lozenge$&\!\!\!\!\!\! \emph{Theory Center, IPNS, KEK, 1-1 Oho, Tsukuba, Ibaraki 305-0801, Japan}\\[.3em]
$^\blacklozenge$&\!\!\!\!\!\! \emph{Graduate University for Advanced Studies (Sokendai), Tsukuba 305-0801, Japan}\\[.3em]
$^\bigstar$&\!\!\!\!\!\!
\emph{Institute for Advanced Research, Nagoya University,
Nagoya 464-8601, Japan}\\[.3em]
$^\bigstar$&\!\!\!\!\!\! \emph{Department of Physics, Nagoya University, Nagoya 464-8602, Japan}\\[.3em]
\end{tabular}

\end{center}
\vskip .6in

\begin{abstract}
\noindent
We develop the stochastic formalism for $\Uone$ gauge fields that has the 
\acl{CS} coupling to a rolling pseudo-scalar field during inflation.
The Langevin equations for the physical electromagnetic fields are derived and the analytic solutions are studied. Using numerical simulation we demonstrate that the electromagnetic fields averaged over the Hubble scale continuously change their direction and their amplitudes fluctuate around the analytically obtained expectation values. 
Though the isotropy is spontaneously broken by picking up a particular local Hubble patch, each Hubble patch is understood independent and the isotropy is conserved globally by 
averaging all the Hubble patches.
\end{abstract}

\end{titlepage}

\renewcommand{\thepage}{\arabic{page}}
\setcounter{page}{1}

\tableofcontents

\section{Introduction}


Inflation gives an elegant explanation for observations of the early universe and is a part of the standard model of cosmology. However, what causes inflation has not been revealed and various models of inflation have been proposed. Among these models, the axion inflation is well motivated, because its shift symmetry ensures the flatness of the inflaton potential which is required for sufficient duration of inflation~\cite{Freese:1990rb,Pajer:2013fsa}. To reheat the universe after inflation, the inflaton
needs to be coupled with other fields. Since the 
\ac{CS} coupling respects the shift symmetry, gauge fields coupled to the inflaton through the CS coupling are often considered and their rich phenomenology has been intensively studied, such as baryogenesis~\cite{Anber:2015yca,Fujita:2016igl,Jimenez:2017cdr,Domcke:2019mnd}, 
leptogenesis~\cite{Domcke:2020quw},
magnetogenesis~\cite{Turner:1987bw, Garretson:1992vt, Anber:2006xt, Fujita:2015iga, Adshead:2016iae, Cuissa:2018oiw}, 
the standard model particle production via the Schwinger effect~\cite{Domcke:2018eki,Domcke:2019qmm,Gorbar:2021rlt,Gorbar:2021zlr,Fujita:2022fwc}, 
and the chiral gravitational wave production~\cite{Cook:2011hg,Barnaby:2011qe,Anber:2012du,Namba:2015gja,Domcke:2016bkh,Adshead:2019igv}.

$\Uone$ gauge field can be generated through the CS coupling during the axion inflation. The mode function of the $\Uone$ gauge field is amplified due to a tachyonic instability, when it exits the Hubble horizon~\cite{Turner:1987bw, Garretson:1992vt, Anber:2006xt}. Although the amplified mode quickly decays on super-horizon scales, a new mode always arises from sub-horizon scales and the gauge field amplitude is persistent. Since each Fourier mode independently evolves and its amplitude is randomly produced from quantum fluctuation, the $\Uone$ gauge field amplitude should fluctuate and its orientation should continuously change in the coordinate space. 
Nevertheless, such stochastic behavior of the gauge field has not been explored in the literature.
This is not just an academic question, rather could be related to phenomenological consequences, such as the resultant baryon asymmetry.
It is desirable to understand whether the stochastic nature of gauge fields could alter the conventional picture.

The stochastic formalism is useful for investigating such stochastic nature of a field caused by quantum fluctuation during inflation (see, e.g., Refs.~\cite{Starobinsky:1982ee,Starobinsky:1986fx,Starobinsky:1994bd} for the first papers), which
is an effective theory for super-horizon fields often called \emph{IR modes}. 
As sub-horizon quantum fluctuations (dubbed \emph{UV modes}) continuously exit the horizon and join the IR modes in the accelerated expansion phase of the universe, the IR \ac{EoM} includes the ``noise'' term as a representative of fluctuations.
In particular, if the UV modes get enhanced around the horizon crossing and can be viewed as classical fields with sufficient squeezing of mode functions, the dynamics of IR modes can be understood as a non-quantum Brownian motion. In this way, one can analyze the behavior of each local horizon patch by means of classical statistical mechanics.
Though the stochastic formalism for scalar fields (both for inflatons and spectators, see, e.g., Ref.~\cite{Pinol:2020cdp} and references therein) has been well established so far, its application to vector fields has not been developed enough, because they are not enhanced by the horizon crossing in their minimal setup. 
The first study of the stochastic formalism for vector fields have addressed a kinetic coupling model where the inflaton is coupled to the kinetic term of $\Uone$ gauge fields~\cite{Fujita:2017lfu}.\footnote{Ref.~\cite{Talebian:2019opf} also studied the stochastic formalism in the kinetic coupling model, but the stochastic equation there does not reproduce the classical background behavior even if the noise term is dropped. This is because the interplay between the inflaton and the gauge field, which enables a classical attractor solution for the gauge field, is not properly taken into account, unlike Ref.~\cite{Fujita:2017lfu}.}
As described above, the \ac{CS} coupling can also source gauge fields in the axion inflation and thus they can be a good target of the stochastic formalism.
However, the previous work on the stochastic formalism of this model claimed that both the electric and magnetic fields were always aligned along ``the $\hat{\bm x}$-direction'' and no rotation of their directions were discussed, which shows a stark contrast to our intuitive argument above~\cite{Talebian:2022jkb}.


In this paper, we develop the stochastic formalism for $\Uone$ gauge fields and explore its implication. We derive the Langevin equation for the $\Uone$ gauge field with the CS coupling to a rolling axion during inflation. The derivation is analogous to that of a scalar field, but has some distinctions. 
We also solve the derived equation to illustrate the stochastic behavior of the $\Uone$ gauge field. 
In particular, our numerical simulation demonstrates that the amplitude fluctuation and the change of the direction based on the above intuitive argument are indeed realized. 

This paper is organized as follows. 
In Sec.~\ref{model}, we briefly explain our setup. 
In Sec.~\ref{Derivation of Langevin equation}, we construct the stochastic formalism for the $\Uone$ gauge field and derive its Langevin equation. 
In Sec.~\ref{Analytic results}, we analytically find the solution and study its properties. 
In Sec.~\ref{Numerical Simulation}, some results of our numerical simulation are shown. 
Sec.~\ref{Conclusion} is devoted to the conclusion of this paper.

\section{Tachyonic growth of gauge fields in the axion inflation}
\label{model}

In this section, we briefly review our model in which the inflaton $\phi$ is coupled to the $\Uone$ gauge field $A_\mu$ through the CS coupling;
\begin{align}
\mathcal{L}=\frac{1}{2}\partial_\mu \phi\partial^\mu \phi-V(\phi)
-\frac{1}{4}F_{\mu\nu}F^{\mu\nu}-\frac{1}{4f}\phi F_{\mu\nu}\tilde{F}^{\mu\nu},
\end{align}
where $F_{\mu\nu}\equiv \partial_\mu A_\nu - \partial_\nu A_\mu$ is the electromagnetic field strength and $\tilde{F}^{\mu\nu}\equiv \epsilon^{\mu\nu\rho\sigma}F_{\rho\sigma}/(2\sqrt{-g})$ is its dual. The determinant of the spacetime metric is denoted by $g$ and the totally anti-symmetric tensor is defined by $\epsilon^{0123}=1$.
In this paper, we do not specify the inflaton potential $V(\phi)$ or the value of the axion decay constant $f$ but simply assume the homogeneous and constant slow roll of the inflaton, $\partial_t\phi=\text{const.}$, as an input parameter in the approximately \ac{dS} background.
The sign of the spacetime metric is defined as $\dd s^2= a^2(\tau)(\dd \tau^2-\dd \bm{x}^2)$. 
Adopting the Coulomb gauge in vacuum, $A_0=\partial_iA_i=0$,
the \ac{EoM} for the comoving gauge field is given by
\begin{align}
\partial_\tau^2 A_i-\partial_j^2 A_i-\frac{1}{f} (\partial_\tau \phi)
\epsilon_{ijl}\partial_j A_l=0,
\label{Original A EoM}
\end{align}
where the conformal time is denoted by $\tau$
and the rank-$3$ totally anti-symmetric tensor is $\epsilon_{123} = 1$.
The gauge field is decomposed by the circular polarization and quantized as
\begin{align}
 A_i(\tau, \bm{x})
 &=
 \sum_{\lambda=\pm} \int \frac{{\rm d}^3 k}{(2\pi)^3}
 e^{i \bm{k \cdot x}} e_{i}^{(\lambda)}(\hat{\bm{k}})
 \hat{A}_\lambda(\tau,\bm k),
 \\
 \hat{A}_\lambda(\tau,\bm k)
 &= \hat{a}_{\bm{k}}^{(\lambda)} \mcA_\lambda(\tau,k)  + \hat{a}_{-\bm{k}}^{(\lambda) \dag} \mcA_\lambda^*(\tau,k),
\label{quantization}
\end{align}
where $e^{(\pm)}_i(\hat{\bm{k}})$ are the right/left-handed polarization vectors defined by $i \bm{k} \cp \bm e^{(\pm)}(\hat{\bm{k}})=\pm k\, \bm{e}^{(\pm)}(\hat{\bm{k}})$, and
$\hat{a}_{\bm{k}}^{(\pm) \dag}$/$\hat{a}_{\bm{k}}^{(\pm)}$ are the creation/annihilation operators which satisfy the commutation relation of 
$[\hat{a}^{(\lambda)}_{\bm{k}},\hat{a}^{(\sigma) \dag}_{-\bm{k}'}] = (2\pi)^3\delta(\bm{k}+\bm{k}')\delta^{\lambda \sigma}$.

During inflation $aH=-1/\tau$, the EoM for the mode function is written as
\begin{align}
\left[ \partial_\tau^2 +k^2 \pm 2k \frac{\xi}{\tau} \right] \mcA_\pm(\tau,k)=0,
\label{EoMforA}
\end{align}
with a characteristic parameter
\begin{align}
\xi\equiv \frac{\partial_\tau \phi}{2f aH}=\frac{\dot{\phi}}{2fH},
\end{align}
where dot denotes the cosmic time derivative.
If $\xi>0$, for instance, $\mcA_+$ modes undergo an exponential enhancement around the horizon crossing, while $\mcA_-$ modes do otherwise.
In the rest of this paper, we take $\xi > 0$ since the solution for $\xi < 0$ is readily obtained by performing the $CP$ transformation to the solution of $\xi > 0$.
With the Bunch--Davies vacuum initial condition and constant $\xi$, one can find the analytic solution for $\mcA_+$ as
\begin{align}
\mcA_+(\tau,k)=\frac{1}{\sqrt{2k}}e^{\pi\xi/2} W_{-i\xi,1/2}(2ik\tau),
\label{A_sol}
\end{align}
where $W_{\alpha,\beta}(z)$ is the Whittaker $W$ function.
This solution approaches a constant asymptotic value in the super-horizon limit,
\begin{align}
\mcA_+(\tau,k) 
\xrightarrow{|k\tau|\ll \xi^{-1}} 
\frac{1}{\sqrt{2k}}\frac{e^{\pi \xi/2}}{\Gamma(1+i\xi)},
\end{align}
where $\Gamma(z)$ is the Gamma function.

\begin{figure}[tbp]
  \includegraphics[width=80mm]{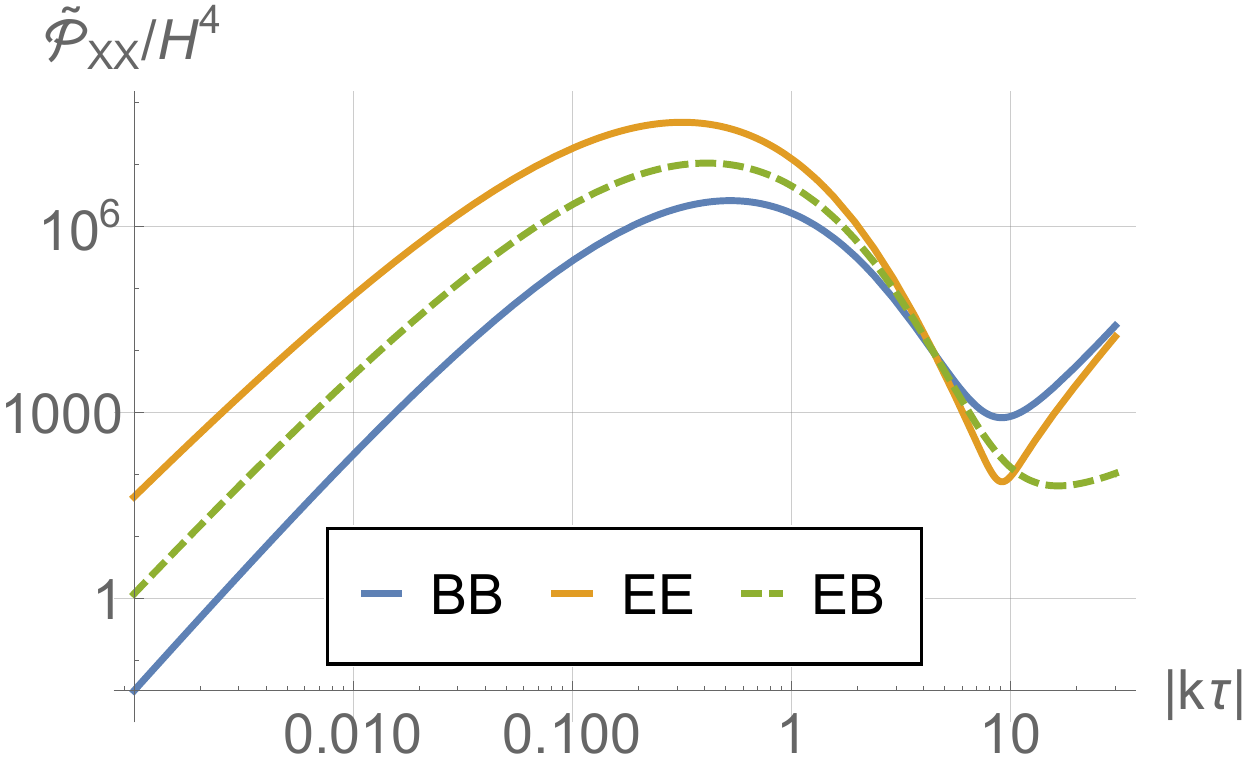}
  \hspace{8mm}
  \includegraphics[width=80mm]{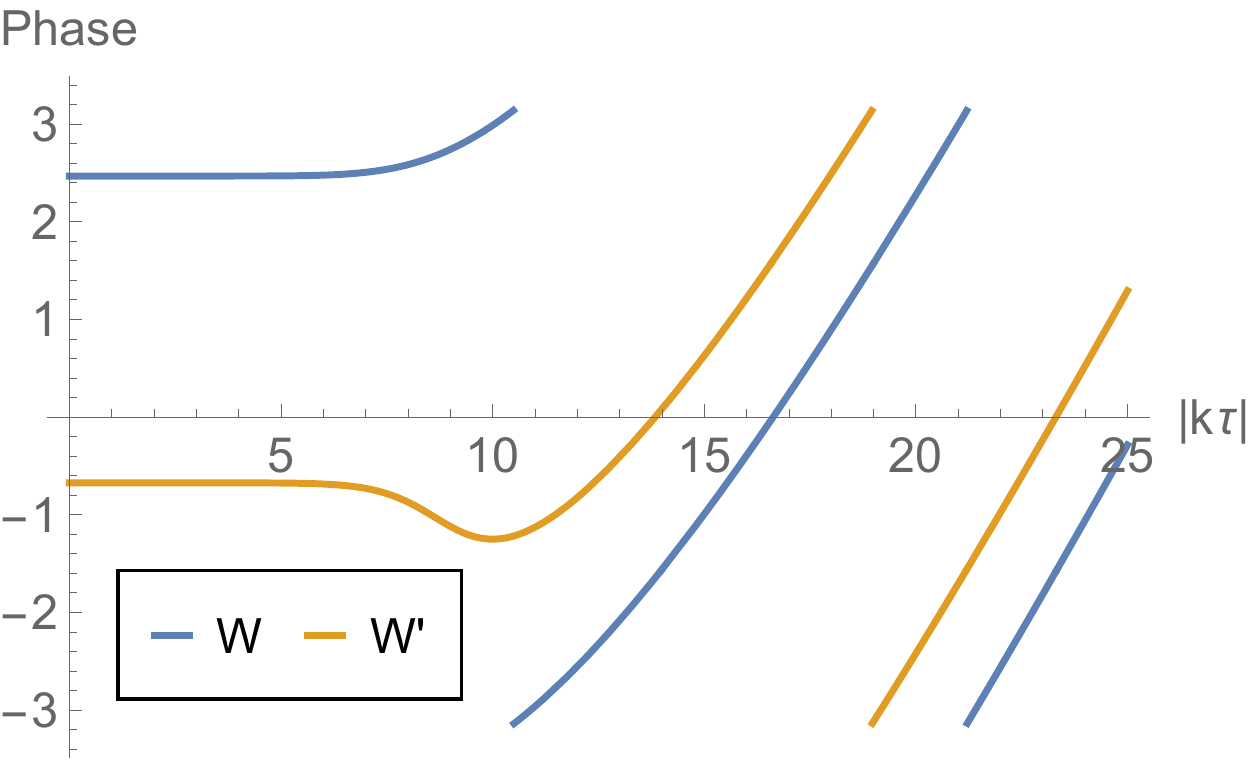}
  \caption
 {{\it (Left panel)} The {\it physical} power spectra $H^{-4}\tilde{\mcP}_{BB}^+$ (blue), $H^{-4}\tilde{\mcP}_{EE}^+$ (orange) and $-H^{-4}(\tilde{\mcP}_{EB}^+ +\tilde{\mcP}_{BE}^+)/2$ (green dashed)
 given in Eqs.\eqref{PB tilde}--\eqref{PBE tilde} for $\xi=5$.
 $\tilde{\mcP}_{EE}^+$ is larger than $|\mathrm{Re}[\tilde{\mcP}_{EB}^+]|$ and $\tilde{\mcP}_{BB}^+$ by $\mathcal{O}(\xi)$ and $\mathcal{O}(\xi^2)$, respectively.
 {\it (Right panel)} The phase of $W$ (blue) and $W'$ (orange) for $\xi=5$.
 The complex phases stop rotating at around $|k\tau|=2\xi$ and these terminal phases are different
 by $\pi$.
 }
 \label{x510}
\end{figure}

With the solution~\eqref{A_sol}, the {\it physical} electromagnetic spectra for the $+$ mode are obtained as
\begin{align}
\tilde{\mcP}_{BB}^+(\tau,k)&=
a^{-4}\mcP_{BB}^+(\tau,k)=
\frac{k^5}{2\pi^2 a^4}\left|\mcA_+(\tau,k)\right|^{2}
=\frac{|k\tau|^4 H^4}{4\pi^2}e^{\pi\xi} \left|W(-k\tau)\right|^2,
\label{PB tilde}
\\
\tilde{\mcP}_{EE}^+(\tau,k)&=
a^{-4}\mcP_{EE}^+(\tau,k) = \frac{k^3}{2\pi^2 a^4} \left|\partial_\tau \mcA_+(\tau,k)\right|^{2}
=\frac{|k\tau|^4 H^4}{4\pi^2 }e^{\pi\xi} \left|W'(-k\tau)\right|^2,
\label{PE tilde}
\\
\tilde{\mcP}_{BE}^+(\tau,k)&=
a^{-4}\mcP_{BE}^+(\tau,k) = -\frac{k^4}{2\pi^2 a^4} \mcA_+(\tau,k)\partial_\tau \mcA_+^*(\tau,k)
=\frac{|k\tau|^4 H^4}{4\pi^2 }e^{\pi\xi} W(-k\tau)W'^*(-k\tau),
\label{PBE tilde}
\end{align}
where $\mcP_{XX}^\lambda$ are the {\it comoving} spectra and $\mcP_{EB}^\lambda=(\mcP_{BE}^{\lambda})^*$.
Here, for brevity, we define the Whittaker function and its derivative as
\begin{align}
W(z) \equiv W_{-i\xi,1/2}(-2i z),
\qquad
W'(z)\equiv \partial_z W_{-i\xi,1/2}(-2iz).
\end{align}
In the left panel of Fig.~\ref{x510}, we present these {\it physical} power spectra. One observes that the spectra reach their peaks at around $|k_\up\tau|\simeq \xi^{-1}$.
The right panel of Fig.~\ref{x510} shows the time evolution of the complex phases of the mode function and its derivative. After the phase rotation stops at $\kappa\simeq 2\xi$, one can treat $\hat{A}_\lambda$ as a classical perturbation~\cite{Polarski:1995jg}.

%
%


\section{Derivation of Langevin equation}
\label{Derivation of Langevin equation}

In this section, we develop the stochastic formalism of the $\Uone$ gauge field.
To go to the stochastic picture, we divide the vector potential $A_i (\tau, \bm x)$ into the IR part and the UV  part,
\begin{align}
\bm A(\tau,\bm{x})=
\sum_{\lambda=\pm}\left[ \bm A_{\rm IR}^\lambda(\tau,\bm{x}) + \bm A_{\rm UV}^\lambda(\tau,\bm{x})\right],
\label{UVIR dec}
\end{align}
with
\begin{align}
\bm A_{\rm IR}^\lambda(\tau,\bm{x}) &\equiv \int \frac{\dd^3 k}{(2\pi)^3} e^{i\bm{k}\cdot\bm{x}}\,\mcW(\tau,k)
\bm e^\lambda(\hat{\bm k})\hat{A}_\lambda(\tau,\bm k),
\label{eq:A_IR}
\\
\bm A_{\rm UV}^\lambda(\tau,\bm{x}) &\equiv \int \frac{\dd^3 k}{(2\pi)^3} e^{i\bm{k}\cdot\bm{x}}\left[1-\mcW(\tau,k)\right]
\bm e^\lambda(\hat{\bm k})\hat{A}_\lambda(\tau,\bm k),
\end{align}
where 
we introduce a window function,
\begin{align}
\mcW(\tau,k)= \Theta(\kappa a H-k).
\end{align}
Here $\Theta(x)$ is the Heaviside function and $\kappa$ is a constant parameter which characterizes the boundary between IR and UV parts. $\bm A_{\rm IR}(
\tau,\bm{x})$ contains only the contributions from the mode functions for $k< \kappa a H$.
We also define the IR and UV parts of its conjugate momentum $\Pi_i(\tau,\bm{x})\equiv A_i'(\tau,\bm{x})$ as
\begin{align}
\bm \Pi_{\rm IR}^\lambda(\tau,\bm{x}) &\equiv \int \frac{\dd^3 k}{(2\pi)^3} e^{i\bm{k}\cdot\bm{x}}\, \mcW(\tau,k)
\bm e^\lambda(\hat{\bm k})\hat{A}'_\lambda(\tau,\bm k),
\\
\bm \Pi_{\rm UV}^\lambda(\tau,\bm{x}) &\equiv \int \frac{\dd^3 k}{(2\pi)^3} e^{i\bm{k}\cdot\bm{x}}\left[1-\mcW(\tau,k)\right]
\bm e^\lambda(\hat{\bm k})\hat{A}'_\lambda(\tau,\bm k).
\end{align}
%
The key point is that due to the time-dependence of the window function, the time derivative of $\bmA_\IR$ does not simply coincide with the IR part of the conjugate momentum $\bm{\Pi}_\IR$ but differs by the mode on the boundary as
\begin{align}
\partial_\tau \bm A_{\rm IR}^\lambda(\tau,\bm{x})-\bm\Pi_{\rm IR}^\lambda(\tau,\bm{x})
= \int \frac{\dd^3 k}{(2\pi)^3} e^{i\bm{k}\cdot\bm{x}}\, \mcW'(\tau,k)
\bm e^\lambda(\hat{\bm k})\hat{A}_\lambda(\tau,\bm k).
\label{A'Pi}
\end{align}
%
Note that the time derivative of the window function yields Dirac's delta function
\begin{align}
\mcW'(\tau,k)=\kappa a^2H^2\delta(\kappa aH-k).
\end{align}
%
Therefore, the \ac{EoM} for the IR modes following the original one~\eqref{Original A EoM} is not closed only by the IR modes but corrected by the transition mode as
\begin{align}
\partial_\tau\bm\Pi_{\rm IR}^\lambda - \bm \nabla^2 \bm A_{\rm IR}^\lambda-\frac{1}{f}\phi'
\curl \bm A_{\rm IR}^\lambda=
\int \frac{\dd^3 k}{(2\pi)^3} e^{i\bm{k}\cdot\bm{x}}\, \mcW'(\tau,k)\bm e^\lambda(\hat{\bm k})\hat{A}'_\lambda(\tau,\bm k),
\end{align}
where the term in the right-hand side represents the new mode joining the IR part.

Here we introduce the IR part of the {\it physical} electromagnetic fields,
%
\begin{align}
\tilde{\bm E}_{\rm IR}^\lambda \equiv - a^{-2}\bm\Pi_{\rm IR}^\lambda,
\qquad
\tilde{\bm B}_{\rm IR}^\lambda \equiv a^{-2}\curl \bm A_{\rm IR}^\lambda.
\label{eq:EB_IR}
\end{align}
Taking the rotation of Eq.~\eqref{A'Pi} and changing the time variable from the conformal time $\tau$ to the cosmic time $t$, 
one finds the stochastic equations for the {\it physical} electromagnetic fields as
\begin{align}
\dot{\tilde{\bm B}}_{\rm IR}^\lambda+2H\tilde{\bm B}_{\rm IR}^\lambda +a^{-1}\curl\tilde{\bm E}_{\rm IR}^\lambda
=\tilde{\bm\Xi}_B^\lambda,
\label{phy B EoM}
\\
\dot{\tilde{\bm E}}_{\rm IR}^\lambda + 2H\tilde{\bm E}_{\rm IR}^\lambda - a^{-1} \curl \tilde{\bm B}_{\rm IR}^\lambda + 2H\xi
\tilde{\bm B}_{\rm IR}^\lambda= \tilde{\bm\Xi}_E^\lambda.
\label{phy E EoM}
\end{align}
%
where we define
\begin{align}
\tilde{\bm\Xi}_B^\lambda(t,\bmx) &\equiv \lambda H\frac{k_\uc (t)}{a^2 (t)}\int \frac{\dd^3 k}{(2\pi)^3} e^{i\bm{k}\cdot\bm{x}}\, \delta(k_\uc(t)-k)
\bm e^\lambda(\hat{\bm k})k \hat{A}_\lambda(\tau,\bm k),
\label{til xi B}
\\
\tilde{\bm\Xi}_E^\lambda(t,\bmx) & \equiv - H \frac{k_\uc (t)}{a^2 (t)}
\int \frac{\dd^3 k}{(2\pi)^3} e^{i\bm{k}\cdot\bm{x}}\, \delta(k_\uc(t)-k)\bm e^\lambda(\hat{\bm k})\hat{A}'_\lambda(\tau,\bm k),
\label{til xi E}
\end{align}
%
with the transition scale $k_\uc(t)\equiv \kappa a(t) H$. 
If one takes a sufficiently small $\kappa\ll2\xi$, these transition modes can be understood as random but classical noise as we saw in the previous section.
Their statistics are inherited from the results of quantum computations as
\bae{
	\braket{\tilde{\Xi}^\lambda_{Xi}(t,\bmx)}=0 \qc 
	\braket{\tilde{\Xi}^\lambda_{Xi}(t,\bmx)\,\tilde{\Xi}^\sigma_{Yj}(t^\prime,\bmy)}=\tilde{\calP}_{XY}^{\lambda}(\kappa)H\delta(t-t^\prime)\delta^{\lambda\sigma}\psi^{\lambda}_{ij}(k_\uc(t)\abs{\bmx-\bmy}),
}
where $X$ and $Y$ denote $B$ or $E$ and we introduced a short-hand notation $\tilde{\mcP}_{BB}^\lambda(\kappa)\equiv \tilde{\mcP}_{BB}^\lambda(\tau,k_\uc(t))$, since it depends only on $-k_\uc(t)\tau =\kappa$ (see Eq.~\eqref{PB tilde}). 
$\psi^\lambda_{ij}(z)$ represents the spherical correlator,
\bae{
	\psi^\pm_{ij}(z)\coloneqq\frac{1}{4\pi}\int\dd{\cos{\theta}}\dd{\phi}e^{iz\cos\theta}e^\pm_i(\hat{\bmk})e^{\pm*}_j(\hat{\bmk})=\pmqty{\frac{z\cos z+(z^2-1)\sin z}{2z^3} & \mp\frac{z\cos z-\sin z}{2z^2} & 0 \\
	\pm\frac{z\cos z-\sin z}{2z^2} & \frac{z\cos z+(z^2-1)\sin z}{2z^3} & 0 \\
	0 & 0 & \frac{-z\cos z+\sin z}{z^3}},
}
with the definition of the polarization vectors $\bme^\pm(\hat{\bmk})=(\cos\varphi\cos\theta\mp i\sin\varphi,\sin\varphi\cos\theta\pm i\cos\varphi,-\sin\theta)^T/\sqrt{2}$ for $\hat{\bmk}=(\sin\theta\cos\varphi,\sin\theta\sin\varphi,\cos\theta)^T$.
As we are interested in the coarse-grained fields, such an oscillating and decaying correlator can be approximated by zero for $z\gg 1$
and by the limit of $z=0$ for $z\ll 1$. 
Although the following discussions do not strongly depend on its intermediate behavior, for simplicity, one can bridge the asymptotic forms with a step function as (see Ref.~\cite{Starobinsky:1994bd})
\bae{
	\psi^\pm_{ij}(z)\simeq\psi^\pm_{ij}(0)\Theta(1-z)=\frac{1}{3}\delta_{ij}\Theta(1-z).
}
This implies that $\tilde{\bm{\Xi}}^\lambda_X$ is understood as a patch-independent ($\propto\Theta(1-k_\uc\abs{\bmx-\bmy})$) and white ($\propto\delta(t-t^\prime)$) Gaussian noise.
Furthermore, if one takes a sufficiently small $\kappa\ll1/\xi$, the gradient terms in the stochastic EoMs can be dropped, 
which is confirmed in the next section.
Under this condition, therefore, the stochastic equation is independent for each local patch, since $\tilde{\bm{\Xi}}^\lambda_X$ is not correlated among patches and the influence from the neighboring patches is negligible.
We hereafter focus on the one-patch dynamics, suppressing the spatial index $\bmx$.

Note that with 
the conformal time and the comoving electromagnetic fields, the noise term would be $\bm{\Xi}_X=a^3 \tilde{\bm \Xi}_X$ and their variances increase in time, $\left< \Xi_{Xi}^\lambda(\tau) \Xi_{Yj}^\sigma(\tau')\right> \propto a^5\delta(\tau'-\tau)$.
Such noise terms are tricky to treat in numerical calculations.  Thus, it is more convenient to handle the {\it physical} electromagnetic fields for which the noise terms have constant variances.
%
%
In the stochastic equations, two polarization modes are decoupled.
Hereafter, we focus on the exponentially amplified mode $\lambda=+$ and suppress the polarization label.

Although we have two noise terms, $\tilde{\bm \Xi}_B$ and $\tilde{\bm \Xi}_E$, they are not independent of each other.
We define a matrix
\begin{align}
\mathcal{M}\equiv 
\frac{4\pi^2}{\kappa^4 H^4 e^{\pi\xi}}
\begin{pmatrix}\tilde{\mcP}_{BB} & \tilde{\mcP}_{BE}\\
\tilde{\mcP}_{EB} & \tilde{\mcP}_{EE} \\
\end{pmatrix}
=
\begin{pmatrix}|W|^2 &  W W'^{*} \\
W^* W' & |W'|^2 \\
\end{pmatrix},
\label{Matrix M}
\end{align}
where all arguments are $\kappa$.
The determinant of this matrix is zero, $\det(\mathcal{M})=0$.
As one observes in the right panel of Fig.~\ref{x510}, for a sufficiently small $\kappa\ll2\xi$, the rotation of the phase of $W$ stops
and $\mathcal{M}$ becomes a real and symmetric matrix,
\begin{align}
\mathcal{M}\xrightarrow{\kappa\ll 2\xi}
\begin{pmatrix}|W|^2 &  -|W| |W'| \\
-|W| |W'| & |W'|^2 \\
\end{pmatrix},
\end{align}
where the off-diagonal part has a minus sign because the phases of $W$
and $W'$ are different by $\pi$ as seen in the right panel of Fig.~\ref{x510}.
Then we can diagonalize it with a rotational matrix
\begin{align}
R=\frac{1}{\sqrt{|W|^2+|W'|^2}}
\begin{pmatrix} |W'| & |W| \\
-|W| & |W'| \\
\end{pmatrix}
\quad\Longrightarrow\quad
R \mathcal{M}R^T=\begin{pmatrix}0 & 0 \\
0 & |W'|^2+|W|^2 \\
\end{pmatrix}.
\label{R def}
\end{align}
Multiplying the stochastic EoMs by this rotational matrix, one finds
\begin{align}\label{eq: rotated EoM}
R\begin{pmatrix}
a^{-2}\partial_t (a^2\tilde{\bm B}_{\rm IR}) \\
a^{-2}\partial_t (a^2 \tilde{\bm E}_{\rm IR})
 + 2H\xi\tilde{\bm B}_{\rm IR} \\
\end{pmatrix}
=
R\begin{pmatrix}
\tilde{\bm\Xi}_B \\
\tilde{\bm\Xi}_E \\
\end{pmatrix}\equiv 
\begin{pmatrix}
\tilde{\bm{\Xi}}_0\\
\tilde{\bm{\Xi}}\\
\end{pmatrix},
\end{align}
where the gradient terms are dropped. $\tilde{\bm{\Xi}}_0\propto |W'| \tilde{\bm\Xi}_B + |W| \tilde{\bm\Xi}_E$ has only vanishing correlations for $\kappa\ll 2\xi$,
\begin{align}
\left< \tilde{\bm \Xi}_{0}\right>=0,
\qquad
\left< \tilde{\Xi}_{0i}(t) \tilde{\Xi}_{0j}(t')\right>
=0,
\qquad
\left< \tilde{\Xi}_{0i}(t) \tilde{\Xi}_{j}(t')\right>
=0.
\end{align}
Thus $\tilde{\bm \Xi}_0$ can be ignored and we have only one noise term $\tilde{\bm \Xi}$.
Note that $\tilde{\calP}_{BE}=\tilde{\calP}_{EB}$ in this limit and thus we hereafter do not distinguish $\braket{\tilde{\bmE}_\IR\cdot\tilde{\bmB}_\IR}$ and $\braket{\tilde{\bmB}_\IR\cdot\tilde{\bmE}_\IR}$.
%
%

The stochastic EoMs now read 
%
\begin{align}\label{eq: stochastic EoM in t}
\begin{pmatrix}
a^{-2}\partial_t (a^2\tilde{\bm B}_{\rm IR}) \\
a^{-2}\partial_t (a^2 \tilde{\bm E}_{\rm IR}) 
 + 2H\xi\tilde{\bm B}_{\rm IR} \\
\end{pmatrix}
\simeq R^T
\begin{pmatrix}
0\\
\tilde{\bm{\Xi}}(t)\\
\end{pmatrix},
\end{align}
where the noise term is characterized by its variance,
\begin{align}
\left<\tilde{\Xi}_i(t) \tilde{\Xi}_j(t')\right>
= \delta_{ij}\delta(t-t')\frac{\kappa^4 H^5}{12\pi^2}e^{\pi\xi}
\left(|W(\kappa)|^2+|W'(\kappa)|^2\right).
\end{align}
%
%
%
This set of equations has a simple interpretation. 
Without the noise terms, the IR electromagnetic fields quickly decay due to the Hubble friction. 
However, thanks to the noise terms
with the constant variance, 
$\tilde{\bm B}_{\rm IR}$ is always sourced by $\tilde{\bm \Xi}_B$
and 
$\tilde{\bm E}_{\rm IR}$ is produced by not only $\tilde{\bm \Xi}_E$
but also $\tilde{\bm B}_{\rm IR}$.

The above equation implies that the noise amplitude significantly depends on $\kappa$. It might look unusual to a reader who is familiar with the stochastic formalism for a massless scalar field, where the noise amplitude is much less sensitive to $\kappa$. Nevertheless, it is not a pathological sign of the formalism. The stochastic formalism is an effective field theory (EFT) for a coarse-grained field and it is normal for an EFT to include its cutoff scale, which is $\Lambda = \kappa aH$ in this case. This $\kappa$ dependence is a consequence of the physical fact that the larger the coarse-grained scale is, the less the averaged fluctuation is. In passing, $\kappa$ should not be too small, for instance, when one computes the backreaction from the gauge field on the inflaton, because the coarse-grained field does not include the most of the power.

\section{Analytic results}
\label{Analytic results}

It is straightforward to obtain the 
formal solutions of Eq.~\eqref{eq: stochastic EoM in t} as
%
\begin{align}
\tilde{\bm B}_{\rm IR}(t)&\simeq a^{-2}(t)\int^t\dd t' a^2(t')\tilde{\bm{\Xi}}_B(t'),
\label{B solution}
\\
\tilde{\bm E}_{\rm IR}(t)&\simeq a^{-2}(t)\left[\int^t_{t_{\rm in}} \dd t' a^2(t')
\tilde{\bm{\Xi}}_E (t') - 2H\xi\int^t\dd t'\int^{t'}\dd t'' a^2(t'')\tilde{\bm{\Xi}}_B(t'')
\right],
\end{align}
where we neglected the initial values of $\tilde{\bm B}_{\rm IR}$
and $\tilde{\bm E}_{\rm IR}$, because their contributions quickly
dilute.
The variance of the IR electromagnetic fields are given by
\begin{align}\label{eq: Bvar}
\left< \tilde{\bm B}_{\rm IR}^2(t)\right>
&\simeq
a^{-4}(t)\iint^t\dd t'\dd t'' a^2(t') a^2(t'')
\left<\tilde{\bm{\Xi}}_B(t')\tilde{\bm{\Xi}}_B(t'')\right>,
\notag\\
&=H\tilde{\mcP}_{BB}(\kappa)\,a^{-4}(t)\int^t\dd t' a^4(t'),
\notag\\
&=\frac{1}{4}\tilde{\mcP}_{BB}(\kappa),
\end{align}
and
\begin{align}\label{eq: Evar}
\left< \tilde{\bm E}_{\rm IR}^2(t)\right>
&\simeq
\frac{1}{4}\tilde{\mcP}_{EE}(\kappa)
- \frac{\xi}{8}\left(\tilde{\mcP}_{BE}(\kappa)+\tilde{\mcP}_{EB}(\kappa)\right)
+\frac{\xi^2}{8}\tilde{\mcP}_{BB}(\kappa).
\end{align}
The cross-correlation is also computed as
\begin{align} \label{eq: EBvar}
\left< \tilde{\bm E}_{\rm IR}(t)\cdot \tilde{\bm B}_{\rm IR}(t)\right>=\frac{1}{4}\tilde{\mcP}_{EB}(\kappa)
- \frac{\xi}{8} \tilde{\mcP}_{BB}(\kappa)\,.    
\end{align}

\begin{figure}[tbp]
  \includegraphics[width=80mm]{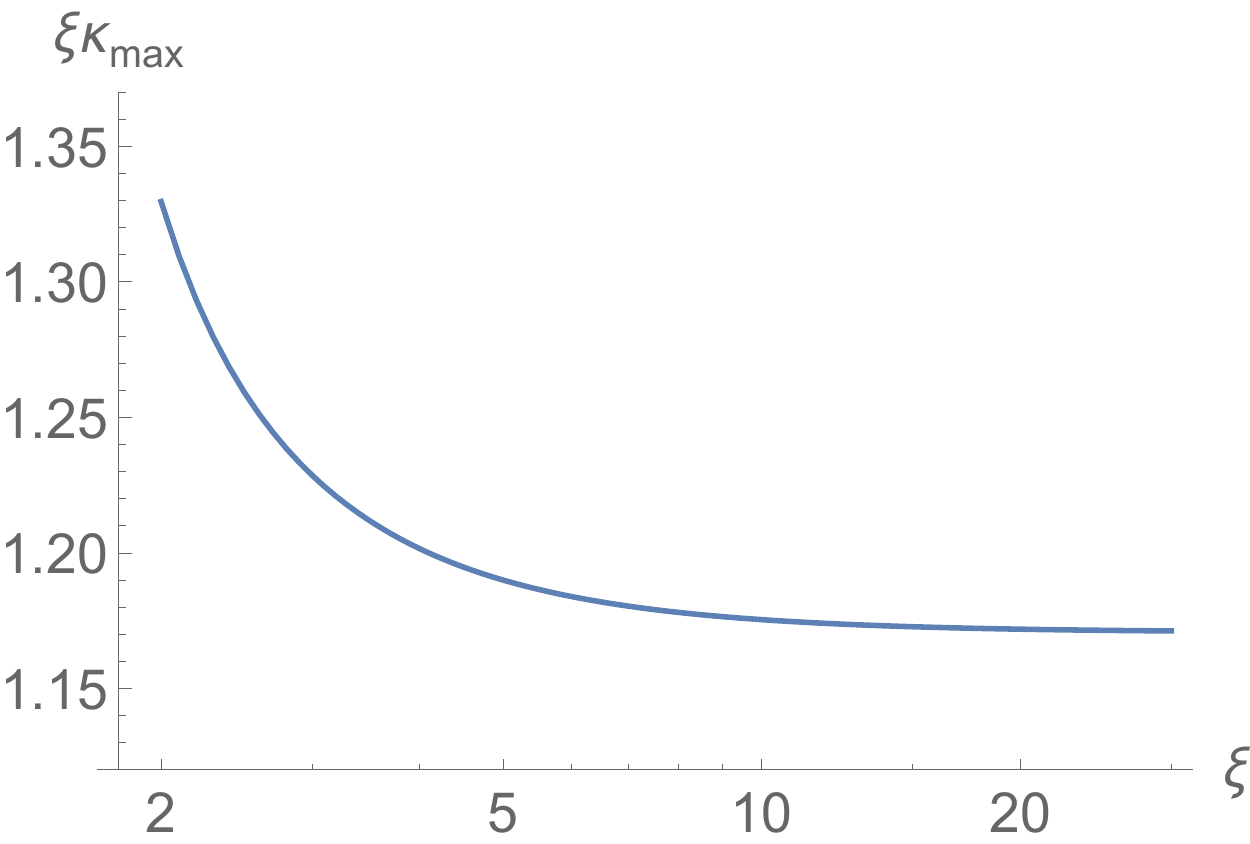}
  \hspace{8mm}
  \includegraphics[width=80mm]{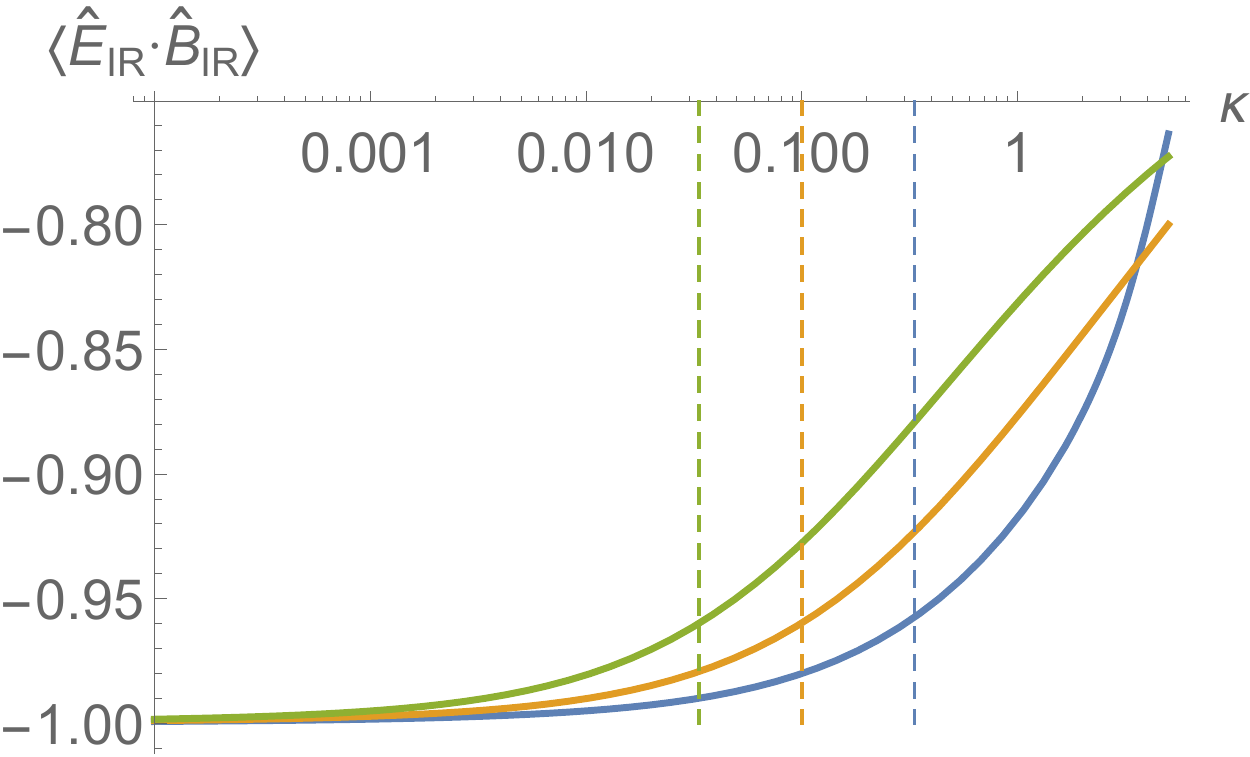}
  \caption
 {{\it (Left panel)} $\kappa_\mathrm{max}$ defined in Eq.~\eqref{saturation condition} multiplied by $\xi$. For $\kappa\ll \kappa_\mathrm{max}\simeq \xi^{-1}$, the gradient term can be safely ignored.
 {\it (Right panel)} $\left< \hat{\bm E}_{\rm IR}\cdot \hat{\bm B}_{\rm IR}\right>$ defined in Eq.~\eqref{eq:EBcorr} against $\kappa$ for $\xi=3$ (blue), $10$ (orange) and $30$ (green). The vertical dashed lines denote  $\epsilon\equiv \xi\kappa=1$ or equivalently $\kappa\simeq\kappa_\mathrm{max}(\xi)$. For $\epsilon\ll 1$, the IR electromagnetic fields are almost completely anti-parallel.
  }
 \label{invest_plot}
\end{figure}

Now we check the consistency of ignoring the gradient terms in the stochastic EoMs.
To ignore the third term compared to the second term
in Eq.~\eqref{phy B EoM},
we need 
%
\begin{align} \label{eq:consistencycheck}
\frac{a^{-1} k_\uc |\tilde{\bm E}_{\rm IR}|}{2H|\tilde{\bm B}_{\rm IR}|}\simeq 
\frac{\kappa}{2}\sqrt{\frac{\left< \tilde{\bm E}_{\rm IR}^2\right>}{\left< \tilde{\bm B}_{\rm IR}^2\right>}}
\ll 1,
\end{align}
where the rotation was evaluated at the cutoff scale, 
$|\curl \tilde{\bm E}_{\rm IR}|\simeq k_\uc|\tilde{\bm E}_{\rm IR}|$.
We introduce $\kappa_\mathrm{max}(\xi)$ which saturates the above condition as
\begin{align}
\frac{\kappa_\mathrm{max}}{2}\sqrt{\frac{\left< \tilde{\bm E}_{\rm IR}^2\right>(\kappa_\mathrm{max},\xi)}{\left< \tilde{\bm B}_{\rm IR}^2\right>(\kappa_\mathrm{max},\xi)}}=1.
\label{saturation condition}
\end{align}
As one can check that it is a monotonically increasing function of $\kappa$, the gradient term can be neglected for $\kappa\ll \kappa_\mathrm{max}$. 
We present numerically computed $\kappa_\mathrm{max}$ in the left panel of Fig.~\ref{invest_plot}.
One observes that $\kappa_\mathrm{max}\simeq 1/\xi$ almost irrespective of the value of $\xi$. Thus the condition to safely neglect the gradient terms is
\begin{align}
\epsilon\equiv \xi\kappa \ll 1.
\label{epsilon condition}
\end{align}
Under this condition, the gradient term in Eq.~\eqref{phy E EoM} can also be ignored  compared with the fourth term, as we focus on a sufficiently large amplification parameter $\xi>1$.
%
%
The condition~\eqref{epsilon condition} is reasonable. Since the {\it physical} power spectra peak at $|k_\up\tau|\sim 1/\xi$ as seen in Fig.~\ref{x510}, we should coarse-grain the electromagnetic fields on a larger scale than the correlation length $|k_\up\tau|^{-1}\sim \xi$ 
to obtain a patch-independent dynamics, which leads to Eq.~\eqref{epsilon condition}.
We also note that, for $\xi > 1$, this gradient condition \eqref{epsilon condition} is tighter than the classicalization condition $\kappa<2\xi$.

Using the analytic solutions~\eqref{eq: Bvar}--\eqref{eq: EBvar}, one finds that the IR electric field and the IR magnetic field are anti-parallel,\footnote{Note that the strict stochastic average of $\hat{\bmE}_\IR\cdot\hat{\bmB}_\IR=\frac{\tilde{\bmE}_\IR\cdot\tilde{\bmB}_\IR}{\sqrt{\tilde{\bmB}_\IR^2\tilde{\bmE}_\IR^2}}$ 
is not equivalent to $\frac{\braket{\tilde{\bmE}_\IR\cdot\tilde{\bmB}_\IR}}{\braket{\tilde{\bmB}_\IR^2}^{1/2}\braket{\tilde{\bmE}_\IR^2}^{1/2}}$. Here we rather call the latter $\braket{\hat{\bmE}_\IR\cdot\hat{\bmB}_\IR}$, which can be calculated analytically and indeed shows the anti-parallelness of the electromagnetic fields in average. Hereafter our discussions do not rely on this definition.}
\begin{align} \label{eq:EBcorr}
\left< \hat{\bm E}_{\rm IR}\cdot \hat{\bm B}_{\rm IR}\right>
\equiv\frac{\left< \tilde{\bm E}_{\rm IR}\cdot \tilde{\bm B}_{\rm IR}\right>}
{\left< \tilde{\bm B}_{\rm IR}^2\right>^{1/2}\left< \tilde{\bm E}_{\rm IR}^2\right>^{1/2}}
\xrightarrow{\epsilon\ll 1} -1.
\end{align}
%
In the right panel of Fig.~\ref{invest_plot}, we present the $\kappa$ dependence of $\left< \hat{\bm E}_{\rm IR}\cdot \hat{\bm B}_{\rm IR}\right>$.
One observes that $\left< \hat{\bm E}_{\rm IR}\cdot \hat{\bm B}_{\rm IR}\right>$ converges to $-1$ for $\epsilon\ll 1$, though a few percents deviation may be found at $\kappa\sim \kappa_\mathrm{max}$.

We also compute the statistical properties of the energy density $\rho_\mathrm{IR}\equiv (\tilde{\bm E}_{\rm IR}^2+\tilde{\bm B}_{\rm IR}^2)/2$ and the inner product $\left< \tilde{\bm E}_{\rm IR}\cdot \tilde{\bm B}_{\rm IR}\right>$ of the IR electromagnetic fields. They appear in the Friedmann equation and the background EoM for the inflaton, respectively, and are of particular interest.
The higher statistical moments of the IR fields are given by 
%
\begin{align}\label{eq: Bkur}
\left< \tilde{\bm B}_{\rm IR}^4\right>
=\frac{5}{3}\left< \tilde{\bm B}_{\rm IR}^2\right>^2,
\qquad
\left< \tilde{\bm E}_{\rm IR}^4\right>
=\frac{5}{3}\left< \tilde{\bm E}_{\rm IR}^2\right>^2,
\qquad
\left< \left(\tilde{\bm E}_{\rm IR}\cdot \tilde{\bm B}_{\rm IR}\right)^2\right>
=\frac{4}{3}\left< \tilde{\bm E}_{\rm IR}\cdot \tilde{\bm B}_{\rm IR}\right>^2
+\frac{1}{3}\left< \tilde{\bm E}_{\rm IR}^2\right>\left< \tilde{\bm B}_{\rm IR}^2\right>.
\end{align}
Note that although a Gaussian scalar random variable $S$ obeys $\left<S^4\right>=3\left<S^2\right>^2$,  3-dimensional vector one $V_i$ with $\left<V_i V_j\right>\propto \delta_{ij}$ 
generally satisfies $\left<V_i V_i V_j V_j\right>=\left<V_i^2\right>\left<V_j^2\right>+2\left<V_i V_j\right>^2=\left<{\bm V}^2\right>^2+(2/3)\left<{\bm V}^2\right>^2=(5/3)\left<{\bm V}^2\right>^2$. 
Using them, one finds the variances normalized by the squared mean value of $\rho_\mathrm{IR}$ and $\left< \tilde{\bm E}_{\rm IR}\cdot \tilde{\bm B}_{\rm IR}\right>$ are 
\begin{align}
&\frac{\left< \rho_\mathrm{IR}^2\right>}{\left< \rho_\mathrm{IR}\right>^2}
=\frac{5\left<\tilde{\bm B}_{\rm IR}^2\right>^2+5\left<\tilde{\bm E}_{\rm IR}^2\right>^2+6\left<\tilde{\bm B}_{\rm IR}^2\right>\left<\tilde{\bm E}_{\rm IR}^2\right>+
4\braket{\tilde{\bmB}_\IR\cdot\tilde{\bmE}_\IR}^2}{3(\left<\tilde{\bm B}_{\rm IR}^2\right>+\left<\tilde{\bm E}_{\rm IR}^2\right>)^2}
\,\xrightarrow{\epsilon\ll 1}\, \frac{5}{3},
\\
&
\frac{\braket{(\tilde{\bmE}_\IR\cdot\tilde{\bmB}_\IR)^2}}{\braket{\tilde{\bmE}_\IR\cdot\tilde{\bmB}_\IR}^2}
=1+
\frac{1}{3}\frac{\left< \tilde{\bm B}_{\rm IR}^2\right>\left< \tilde{\bm E}_{\rm IR}^2\right>+\left<\tilde{\bm B}_{\rm IR}\cdot\tilde{\bm E}_{\rm IR}\right>^2}{
\braket{\tilde{\bmE}_\IR\cdot\tilde{\bmB}_\IR}^2}
\,\xrightarrow{\epsilon\ll 1}\, \frac{5}{3}\,,
\end{align}
where we used Eq.~\eqref{eq:EBcorr} in the first line, and the convergence of the second line is similar to $\left< \hat{\bm E}_{\rm IR}\cdot \hat{\bm B}_{\rm IR}\right>$ shown in the right panel of Fig.~\ref{invest_plot}. Hence, their variances have the same statistics as the kurtosis of a 3-dimensional Gaussian vector variable and are smaller than that of a Gaussian scalar variable. 

Before closing this section, we consider the correlation time of the IR electromagnetic fields. 
Rewriting the solution~\eqref{B solution} into $a^2(t+\Delta t)\tilde{\bm B}_{\rm IR}(t+\Delta t)-a^2(t)\tilde{\bm B}_{\rm IR}(t)=\int^{t+\Delta t}_{t}\dd t' a^2(t')\tilde{\bm{\Xi}}_B(t')$, 
and doing the same for $\tilde{\bm E}_\mathrm{IR}$, one can show
\begin{align}
\left< \tilde{\bm B}_{\rm IR}(t)\cdot
\tilde{\bm B}_{\rm IR}(t+\Delta t) \right>
&=e^{-2H\Delta t} \tilde{\bm B}_{\rm IR}^2 (t),
\label{Bi Be}
\\
\left< \tilde{\bm E}_{\rm IR}(t)\cdot
\tilde{\bm E}_{\rm IR}(t+\Delta t) \right>
&=e^{-2H\Delta t} 
\left[\tilde{\bm E}_{\rm IR}^2(t) - 2\xi H\Delta t\,
\tilde{\bm E}_{\rm IR}(t)\cdot\tilde{\bm B}_{\rm IR}(t)\right].
\label{Ei Ee}
\end{align}
%
Although the second term in Eq.~\eqref{Ei Ee} gives a linear correction, the both correlations exponentially decay. The characteristic time scale is
\begin{align}
t_\uc=\frac{1}{2H}.
\label{correlation time}
\end{align}
Therefore, the IR electromagnetic fields take new independent values every half Hubble time.

\section{Numerical Simulation}
\label{Numerical Simulation}

In this section we numerically simulate the IR electromagnetic fields and illustrate their behaviors. 
To make variables dimensionless, the time variable is often normalized by the Hubble parameter, that is, we use the
e-folding number $N=\int^t_0H\dd{t^\prime}$ 
as a time variable. 
The stochastic EoM~\eqref{eq: stochastic EoM in t} can be rewritten in $N$ as
\bae{\label{num EoM}
    \pmqty{
        \partial_N\tilde{\bm{B}}_\mathrm{IR}+2\tilde{\bm{B}}_\mathrm{IR} \\
        \partial_N\tilde{\bm{E}}_\mathrm{IR}+2\tilde{\bm{E}}_\mathrm{IR} + 2\xi\tilde{\bm{B}}_\mathrm{IR}
    }=R^T
    \pmqty{
        0 \\
        \tilde{\bm{\Xi}}(N)
    },
}
with
\bae{
    \braket{\tilde{\Xi}_i(N)\tilde{\Xi}_j(N^\prime)}=\delta_{ij}\delta(N-N^\prime)\frac{\kappa^4H^4}{12\pi^2}e^{\pi\xi}(|W|^2+|W^\prime|^2).
}

\begin{figure}
    \centering
    \begin{tabular}{c}
        \begin{minipage}{0.5\hsize}
            \centering
            \includegraphics[width=0.95\hsize]{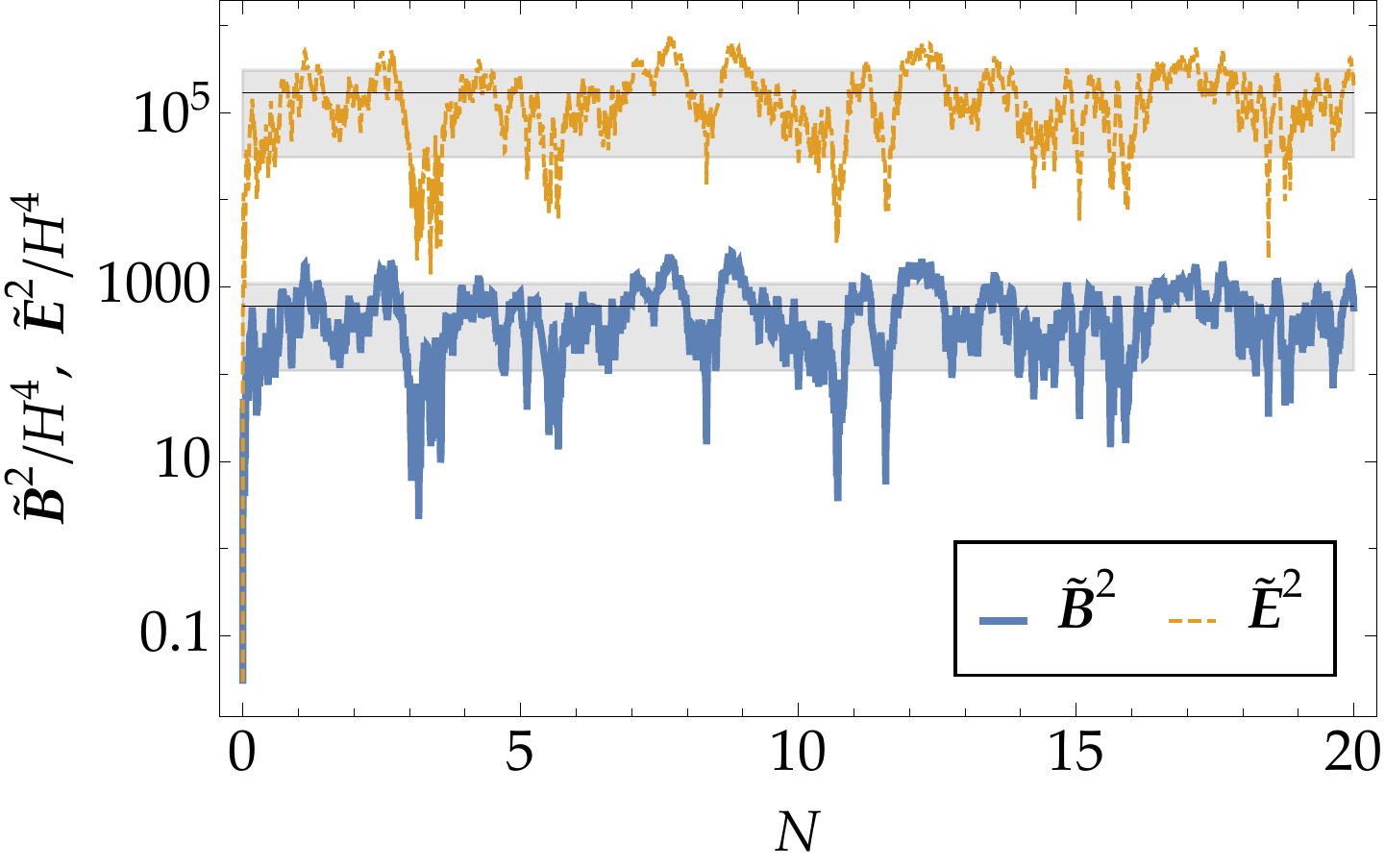}
        \end{minipage}
        \begin{minipage}{0.5\hsize}
            \centering
            \includegraphics[width=0.95\hsize]{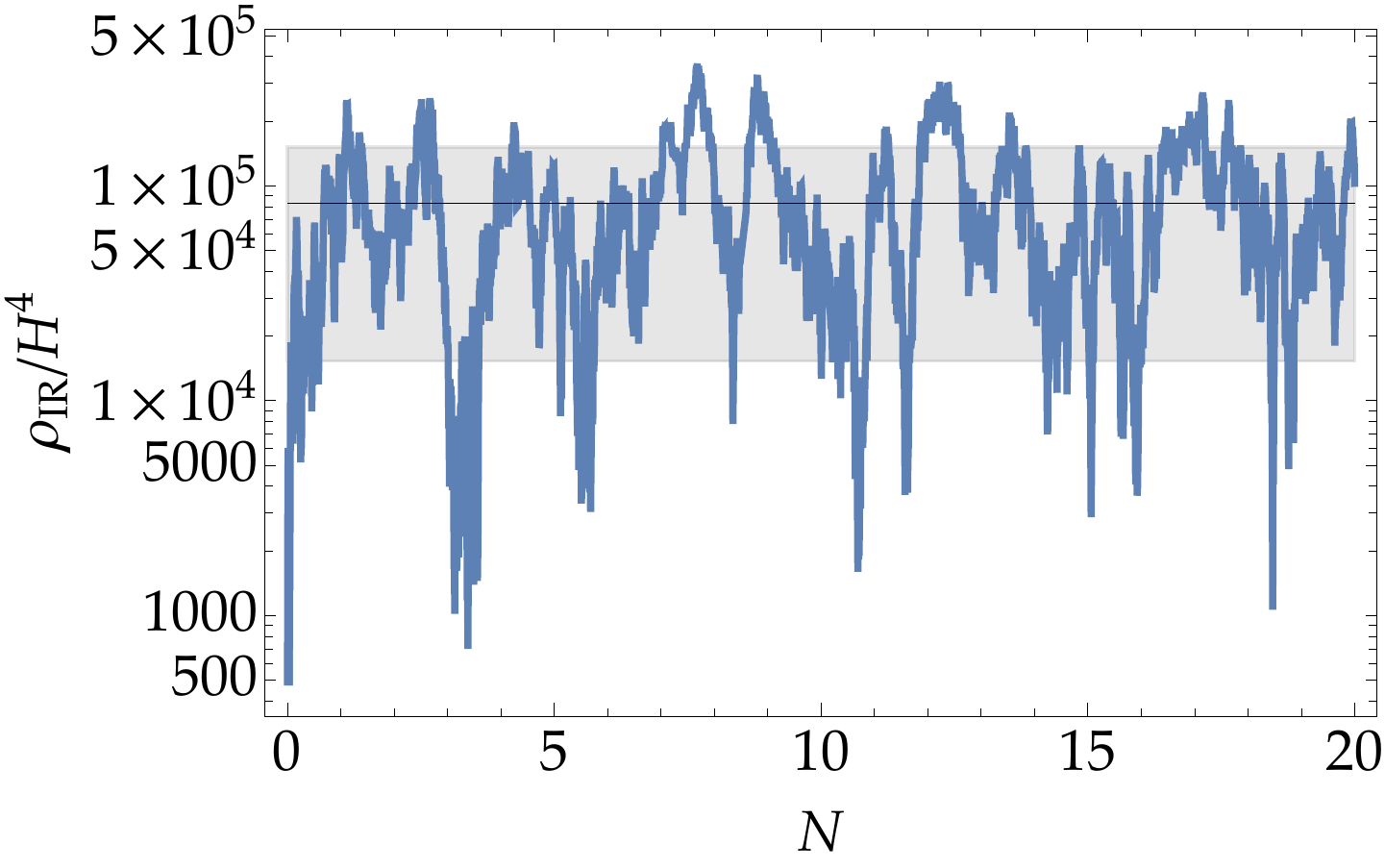}
        \end{minipage} \\
        \begin{minipage}{0.5\hsize}
            \centering
            \includegraphics[width=0.95\hsize]{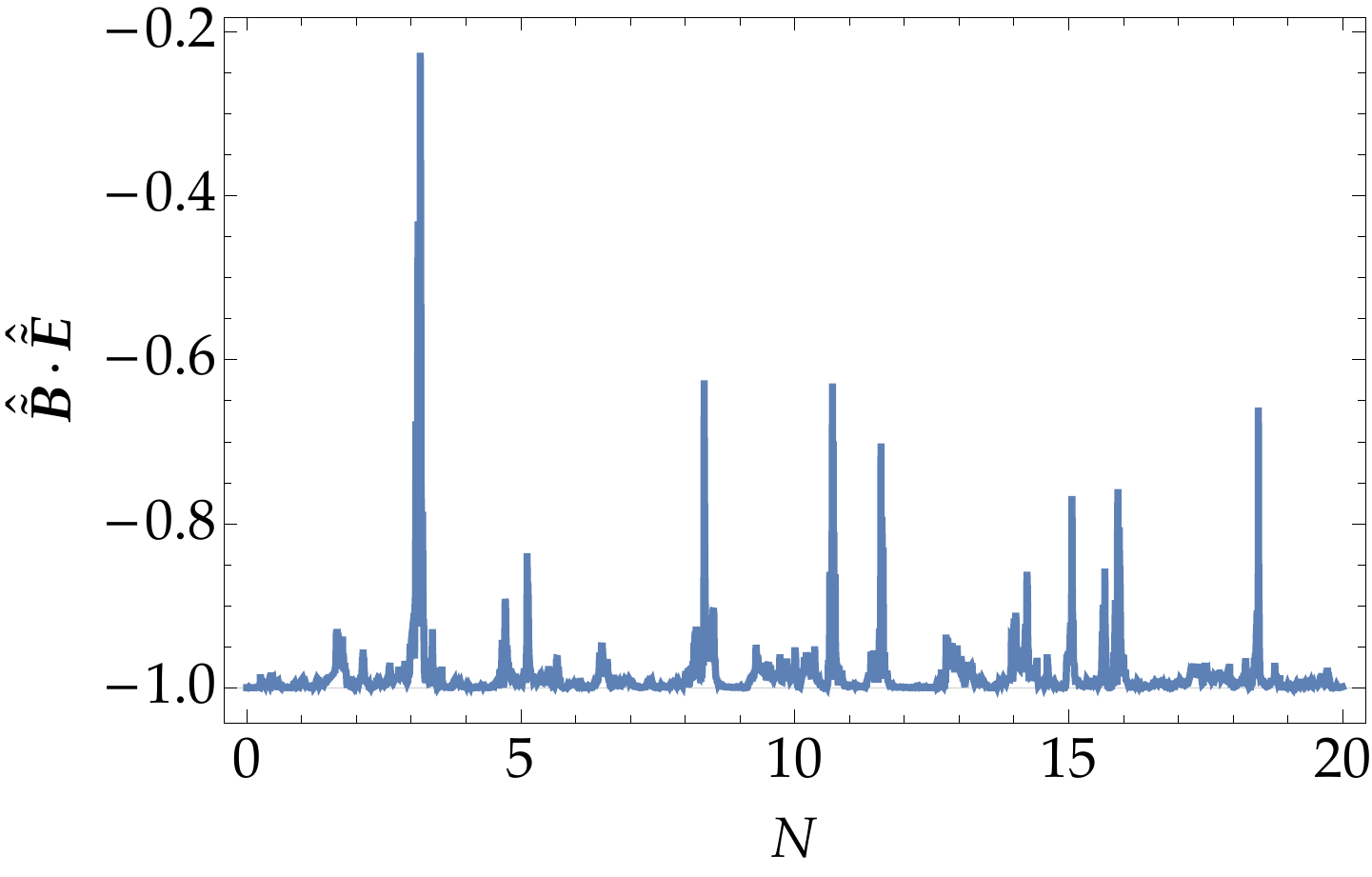}
        \end{minipage} 
    \end{tabular}
    \caption{One realization of the IR electromagnetic fields numerically computed based on Eq.~\eqref{num EoM} with the parameters $H=10^{-5}\Mpl$, $\xi=5$, and $\kappa=0.02$, and vanishing electromagnetic fields at initial time $N=0$. 
    {(\it Top-Left panel)} The amplitudes $\tilde{\bm{B}}^2_\mathrm{IR}$ (blue) and $\tilde{\bm{E}}^2_\mathrm{IR}$ (orange dashed) normalized by 
    $H^4$ against the e-folding number $N$. From bottom to top, the horizontal 
    black thin lines show the analytic estimations of the mean amplitudes~\eqref{eq: Bvar} and \eqref{eq: Evar} and gray bands indicate their standard deviations~\eqref{eq: std deviation}. 
    \emph{(Top-Right panel)} The similar plot for $\rho_\IR=(\tilde{\bmE}_\IR^2+\tilde{\bmB}_\IR^2)/2$.
    {\it (
    Bottom panel)} The normalized inner product of the IR electromagnetic fields, 
    $\hat{\bm{B}}_\mathrm{IR}\cdot\hat{\bm{E}}_\mathrm{IR}\equiv \tilde{\bm{B}}_\mathrm{IR}\cdot\tilde{\bm{E}}_\mathrm{IR}/(|\tilde{\bm{B}}_\mathrm{IR}||\tilde{\bm{E}}_\mathrm{IR}|)$. 
    It stochastically fluctuates around $-1$. 
    }
    \label{fig: sim}
\end{figure}

In Fig.~\ref{fig: sim}, we present the amplitudes, the energy density, and the inner product of the IR electromagnetic fields in one realization of our numerical simulations, starting from the vanishing field value at the initial time $N=0$.
The analytically estimated mean values~\eqref{eq: Bvar} and \eqref{eq: Evar} and their standard deviations
\bae{\label{eq: std deviation}
    \sqrt{\braket{(\tilde{\bmB}_\IR^2)^2}-\braket{\tilde{\bmB}_\IR^2}^2}=\sqrt{\frac{2}{3}}\braket{\tilde{\bmB}_\IR^2} \qc
    \sqrt{\braket{(\tilde{\bmE}_\IR^2)^2}-\braket{\tilde{\bmE}_\IR^2}^2}=\sqrt{\frac{2}{3}}\braket{\tilde{\bmE}_\IR^2},
}
derived from Eq.~\eqref{eq: Bkur} are also shown.
One finds that the amplitudes of $\tilde{B}_\mathrm{IR}$ and $\tilde{E}_\mathrm{IR}$ shown in the top-left panel rapidly reach and stay around the predicted averages within the estimated errors, which indicates the superhorizon electric/magnetic fields are dominated by the stochastic noise.
It is interesting to note that these two amplitudes fluctuate in a very similar way,
because they are sourced by the same noise $\tilde{\bm{\Xi}}$.
The similar plot for the energy density $\rho_\IR=(\tilde{\bmE}_\IR^2+\tilde{\bmB}_\IR^2)/2$ is shown in the top-right panel.
The bottom panel 
confirms that the unit vectors 
$\hat{\bm{B}}_\mathrm{IR}=\tilde{\bm{B}}_\mathrm{IR}/|\tilde{\bm{B}}_\mathrm{IR}|$ 
and $\hat{\bm{E}}_\mathrm{IR}=\tilde{\bm{E}}_\mathrm{IR}/|\tilde{\bm{E}}_\mathrm{IR}|$ are in the anti-parallel configuration $\hat{\bm{B}}_\mathrm{IR}\cdot\hat{\bm{E}}_\mathrm{IR} = -1$ for the most of time. 

\begin{figure}
    \centering
    \begin{tabular}{c}
        \begin{minipage}{0.5\hsize}
            \centering
            \includegraphics[width=0.8\hsize]{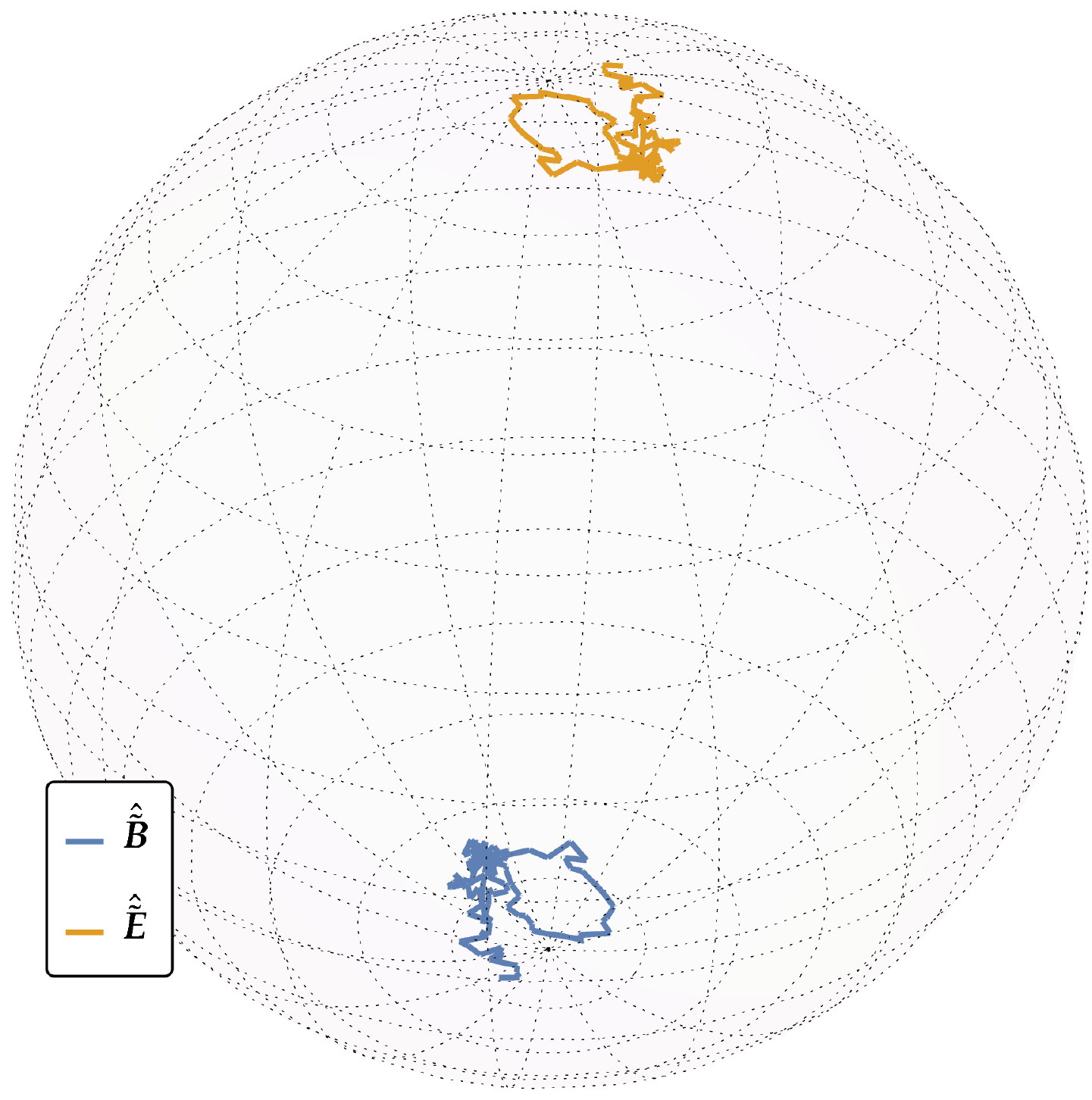}
        \end{minipage}
        \begin{minipage}{0.5\hsize}
            \centering
            \includegraphics[width=0.8\hsize]{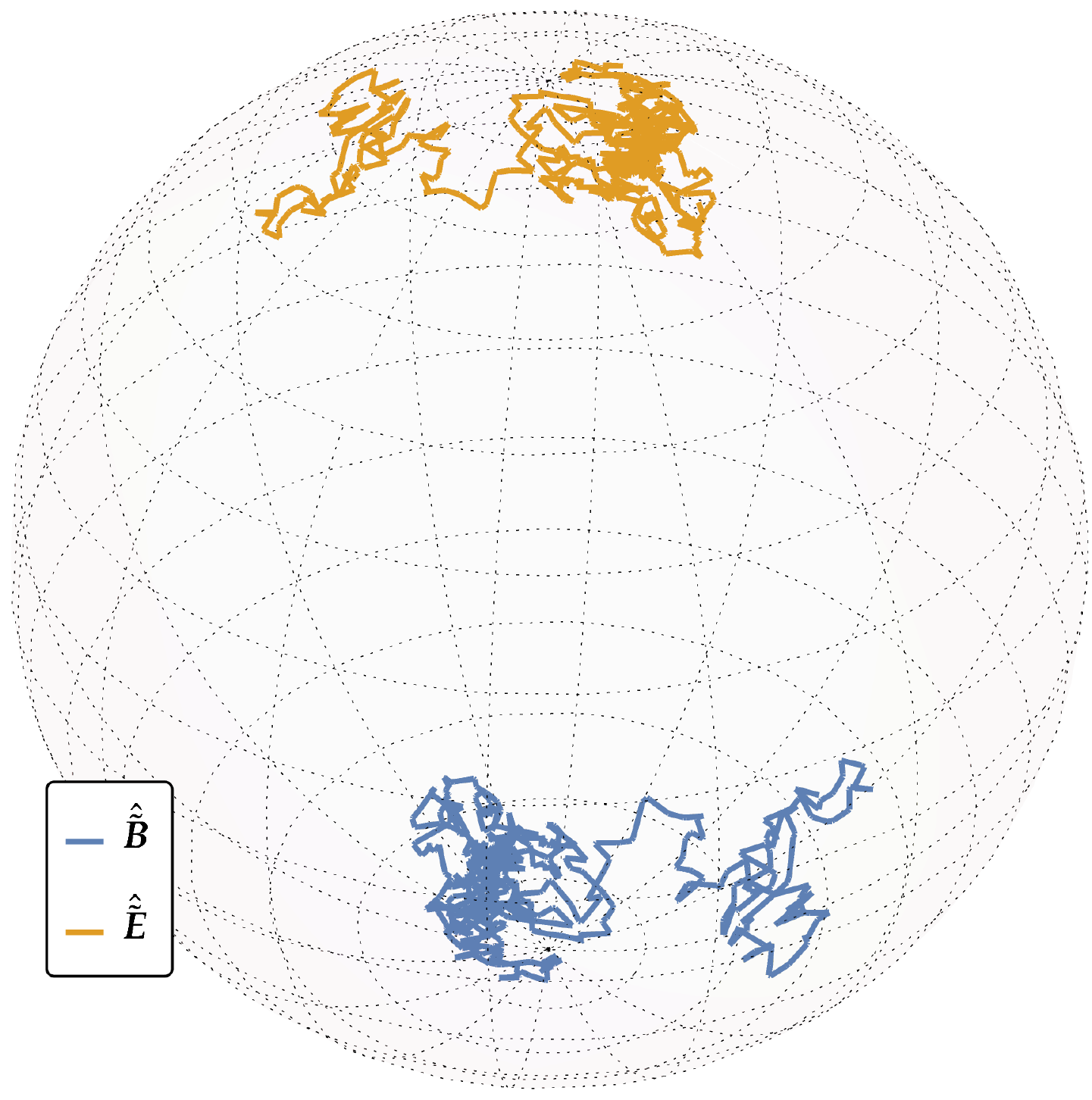}
        \end{minipage}
        \\\\
        \begin{minipage}{0.5\hsize}
            \centering
            \includegraphics[width=0.8\hsize]{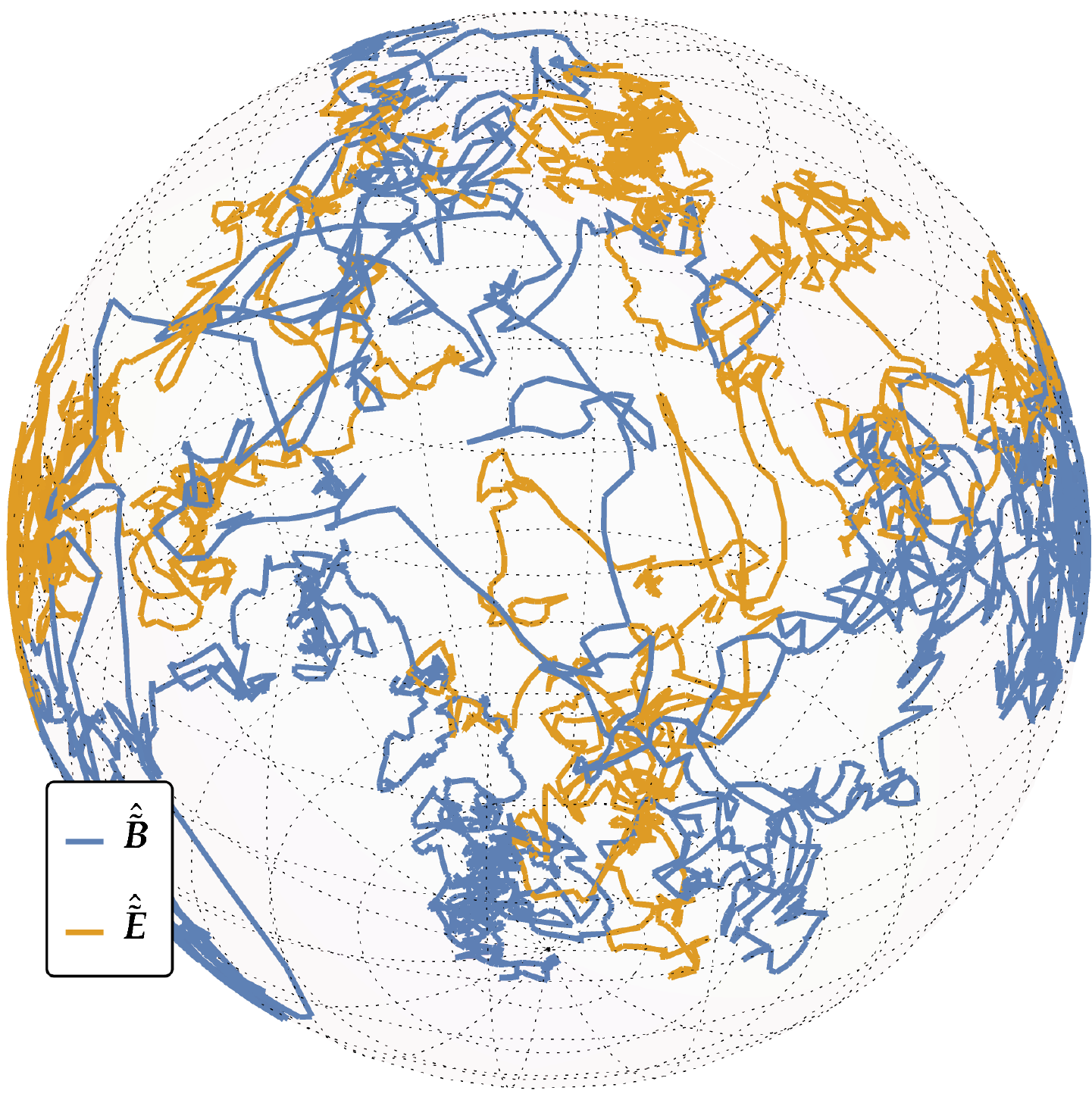}
        \end{minipage}
        \begin{minipage}{0.5\hsize}
            \centering
            \includegraphics[width=0.8\hsize]{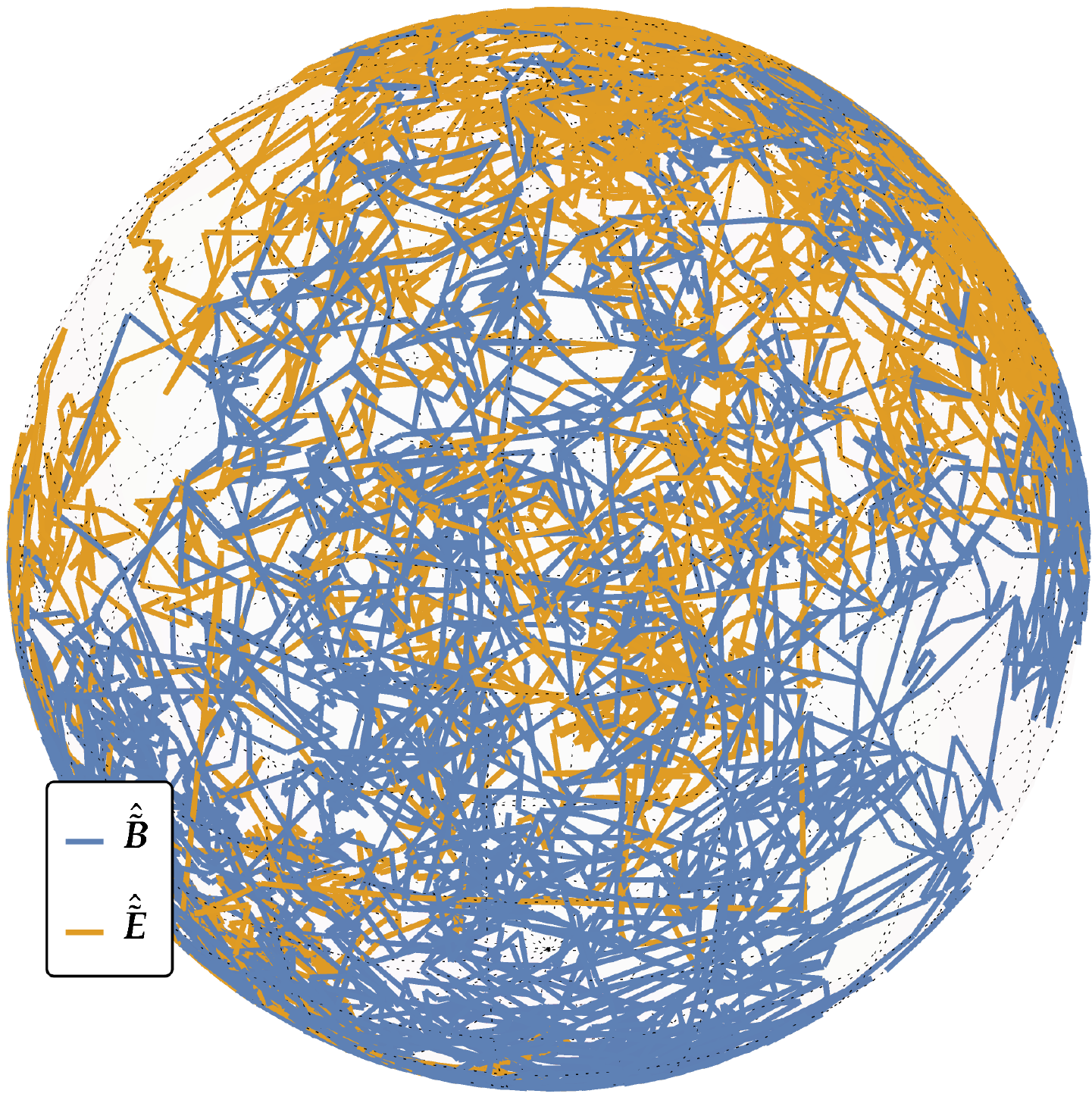}
        \end{minipage}
        \end{tabular}
    \caption{The 3-dimensional trajectories of the unit vectors $\hat{\bm{B}}$ (blue) and $\hat{\bm{E}}$ (orange) in 
    the same realization 
    as Fig.~\ref{fig: sim}. 
    These panels show them for the duration $\Delta N$ of $0.1$ (top-left), $0.5$ (top-right), $2$ (bottom-left), and $10$ e-folds (bottom-right). 
    $\hat{\bm{B}}$ and $\hat{\bm{E}}$ are anti-parallel and do not significantly change the direction for $\Delta N\lesssim 0.5$ as expected from  Eqs.~\eqref{eq:EBcorr} and \eqref{correlation time}. For a longer time scale, however, they randomly take other directions and eventually sweep all directions, 
    keeping $\hat{\bm{B}}\cdot\hat{\bm{E}}\simeq -1$.}
    \label{fig: sphere}
\end{figure}
In Fig.~\ref{fig: sphere}, we present the representative trajectories of the unit vectors, $\hat{\bm{B}}_\mathrm{IR}$ and $\hat{\bm{E}}_\mathrm{IR}$.
Within the correlation time $t_\uc=(2H)^{-1}$ or $N_\uc=1/2$, they do not significantly change the direction.
However, since they lose their memories of the past directions over the correlation time $t \gtrsim t_\uc$, they are oriented in random directions and the trajectories finally sweep the entire 3-dimensional sphere.
This result demonstrates that the IR electromagnetic fields continually change their directions during inflation and analysis under the approximation of static electromagnetic fields may fail to capture its interesting consequences in the present model.


\section{Conclusion}
\label{Conclusion}

In this paper, we developed the stochastic formalism of $\Uone$ gauge fields coupled to a rolling pseudo-scalar field during inflation.
The derivation of the stochastic (Langevin) EoMs for $\Uone$ gauge fields is analogous to that for a scalar field, while we had the following two features. First, the variances of the noise terms become constant for the {\it physical} electromagnetic fields, $\tilde{\bm E}\propto a^{-2}\bm E$ and $\tilde{\bm B}\propto a^{-2}\bm B$, in the cosmic time $t$ in the \ac{dS} limit. 
Second, although two different noise terms $\tilde{\bm \Xi}_E$ and $\tilde{\bm \Xi}_B$ appeared in the course of the derivation, we diagonalized them and found only one vector noise term $\tilde{\bm \Xi}$ is relevant for the IR modes. Thus, one needs a single 3-dimensional Gaussian random variable to compute the behaviors of the electromagnetic fields. This is actually in the same situation as a standard scalar field case: a noise for a scalar field $\Xi_\phi$ and one for its conjugate momentum $\Xi_\pi$ are caused by a single noise (see, e.g., Ref.~\cite{Pinol:2020cdp}).

We investigated the derived stochastic EoMs in both analytic and numerical ways. We analytically found that the expected values of the electromagnetic amplitudes are constants given by their power spectra, and the electric and magnetic fields are expected to be anti-parallel.  
Moreover, the variance of their energy density is $5/3$ of its mean value squared, which is smaller than the kurtosis of a scalar Gaussian variable because more degrees of freedom are involved. Our numerical simulation demonstrated that the electromagnetic fields randomly change their directions over the coherent time scale, while keeping the anti-parallel configuration. Since this continuous change of direction of the electromagnetic fields has not been discussed in the previous works, 
it would be interesting to explore its implication for related phenomenology.
Note that the isotropy is spontaneously broken when we pick up one particular configuration realized in a local Hubble patch. However, each Hubble patch is understood independent and the expectation values are obtained by averaging all the Hubble patches.
As the direction of the gauge field is random for each Hubble patch,
the isotropy is conserved in this sense.

The mode function and power spectra were analytically obtained in Sec.~\ref{model}, and in principle one should be able to read off all physical information from these expressions. Seen in that light, the stochastic analysis does not provide completely new findings. However, we believe it is valuable to have the stochastic formalism for gauge field dynamics for the following two reasons. (i) The stochastic approach offers visible picture of dynamics, with which it is easier to understand what actually happens. In particular, as demonstrated in Figs.~\ref{fig: sim} and \ref{fig: sphere}, visualization enabled by our stochastic formalism is powerful. What we have imagined from the analytic formulae can be clearly presented. (ii) We can further develop the stochastic formalism for extended theories of gauge fields based on this simple example. Some models of SU($N\ge 2$) gauge fields and U(1) gauge fields with charged particles have attracted attention recently in the context of inflation (see, e.g., Refs.~\cite{Adshead:2012kp,Fujita:2021eue} and \cite{Domcke:2018eki,Domcke:2019qmm,Gorbar:2021rlt,Gorbar:2021zlr,Fujita:2022fwc}). Even in the present model, the inflation dynamics can be turned on (i.e. $\xi\neq \text{const.}$). In these cases, the EoMs for the gauge fields are non-linear or coupled, and their analytic solutions are no longer available. We expect that the stochastic analysis will provide new insights in such cases.

Our formalism should be carefully extended to compute the spatial distribution of $\Uone$ gauge fields. If one allocates multiple IR electromagnetic fields in spatial grid positions, which independently evolve based on the stochastic EoMs, Gauss's law $\bm \nabla\cdot\tilde{\bm E}_\mathrm{IR}(t,\bm x)=0$ would be violated. 
Note that the constraint condition coming from the Euler–Lagrange equation for $A_0$ corresponds to Gauss's law in the present case with the temporal gauge $A_0=0$.
Gauss's law is trivially satisfied at the leading order in the gradient expansion, but beyond the leading order, both the gradient terms in the EoM and the spatial correlations of noises should be consistently taken into account.
That is compatible in itself with the stochastic formalism, and one can implement it in principle, though it may complicate the calculation procedure. 
This issue does not matter as long as the IR fields at a single spatial point is computed by neglecting their gradient. 

\section*{Acknowledgments}
We would like to thank Hassan Firouzjahi for useful discussion and Kohei Kamada for kindly letting us know that our first arXiv version lacked acknowledgement.
This work was supported in part by the Japan Society for the Promotion of Science (JSPS) KAKENHI, Grant Number
JP18K13537, JP20H05854 (T.F.), 
JP19K14707, and JP21K13918 (Y.T.).
K.M.\ was supported by MEXT Leading Initiative for Excellent Young Researchers Grant Number JPMXS0320200430.

\bibliographystyle{JHEP}
\bibliography{Stochastic}
\end{document}